# On the interactions of transverse ion-cyclotron waves with ions in solar wind plasma

Sofiane Bourouaine[1], Eckart Marsch[1] and Fritz M. Neubauer[2]

## ABSTRACT

We show the evidence of ion-cyclotron dissipation mechanism in solar wind plasma using Helios data. From our statistical analysis we found that the wave power of high-frequency transverse waves (having frequencies between 0.01 and 1 normalized to the proton gyrofrequency in the plasma frame) correlates with both, the proton temperature anisotropy, $T_\perp/T_\parallel$, and the normalized differential speed, $V_{\alpha p}/V_A$, between alpha particles and protons. Furthermore, when this speed stays below 0.5, then the alpha-particle temperature anisotropy correlate positively with the relative power of the transverse waves. However, if $V_{\alpha p}/V_A$ is larger than 0.6, then the alpha-particle temperature anisotropy tends to decrease towards values below unity, despite the presence of transverse waves with relatively large amplitudes. For small relative wave amplitude, it is found that alpha particles can even be heated more strongly than protons when the alpha-to-electron density ratio nearly or below 0.01. Our findings are in good agreement with predictions of kinetic theory for the resonant interaction of ions with Alfvn-cyclotron waves and for the resulting wave dissipation. Also, the study suggests that the turbulence could lead to the generation of parallel Alfven cyclotron waves.

[1]Max-Planck-Institut für Sonnensystemforschung, 37191 Katlenburg-Lindau, Germany

[2]Institut für Geophysik und Meteorologie, Universität zu Köln, Albertus-Magnus-Platz, Köln, 50923, Germany